# Monolithic Infrared Silicon Photonics: The Rise of (Si)GeSn Semiconductors

O. Moutanabbir,[1,*] S. Assali,[1] X. Gong,[2] E. O'Reilly,[3] C. Broderick,[3] B. Marzban,[4] J. Witzens,[4] W. Du,[5] S-Q. Yu,[6] A. Chelnokov,[7] D. Buca,[8] D. Nam[9]

[1] *Department of Engineering Physics, École Polytechnique de Montréal, Québec, Canada*
[2] *Department of Electrical and Computer Engineering National University of Singapore, Singapore*
[3] *Tyndall National Institute, University College Cork, Ireland*
[4] *Institute of Integrated Photonics, RWTH Aachen University, Aachen, Germany*
[5] *Department of Electrical Engineering and Physics, Wilkes University, PA, USA*
[6] *Department of Electrical Engineering, University of Arkansas, AR, USA*
[7] *Univ. Grenoble Alpes, CEA, Leti, Grenoble, France*
[8] *Peter Gruenberg Institute 9 (PGI-9), Forschungszentrum Jülich, Jülich, Germany*
[9] *School of Electrical and Electronic Engineering, Nanyang Technological University, Singapore*
[*] Email: oussama.moutanabbir@polymtl.ca

**Abstract**:

(Si)GeSn semiconductors are finally coming of age after a long gestation period. The demonstration of device-quality epi-layers and quantum-engineered heterostructures has meant that tunable all-group IV Si-integrated infrared photonics is now a real possibility. Notwithstanding the recent exciting developments in (Si)GeSn materials and devices, this family of semiconductors is still facing serious limitations that need to be addressed to enable reliable and scalable applications. The main outstanding challenges include the difficulty to grow high-crystalline quality layers and heterostructures at the desired Sn content and lattice strain, preserve the material integrity during growth and throughout device processing steps, and control doping and defect density. Other challenges are related to the lack of optimized device designs and predictive theoretical models to evaluate and simulate the fundamental properties and performance of (Si)GeSn layers and heterostructures. This Perspective highlights key strategies to circumvent these hurdles and bring this material system to maturity to create far-reaching new opportunities for Si-compatible infrared photodetectors, sensors, and emitters for applications in free-space communication, infrared harvesting, biological and chemical sensing, and thermal imaging.

(Si)GeSn alloys constitute isovalent substitution of the group-IV element Sn in cubic diamond-structured (Si)Ge lattices. This emerging family of semiconductors provides strain and composition as two degrees of freedom to independently engineer the lattice parameter and the band structure, in a similar fashion to the mature compound semiconductors. The prospect of mimicking III-V and II-VI heterostructures and devices using all-group IV semiconductors on a Si platform has generated a great deal of interest, motivated by the potential to achieve the long-sought-after monolithic integration of electronics and photonics.[1-8] Fig. 1(a) illustrates this potential by showing the bandgap energy-lattice parameter space that can be covered by these semiconductors. Short-wave infrared (SWIR: 1.5-3 μm), mid-wave infrared (MWIR: 3-8 μm), and long-wave infrared (LWIR: 8-14 μm) bands can be potentially served by all-group IV (Si)GeSn devices. Free-space data communications, infrared harvesting, biochemical sensing, eye-safe LiDAR applications, and thermography technologies would strongly benefit from the availability of cost-effective and Si-compatible (Si)GeSn opto-electronic devices. Moreover, the bandgap is expected to progressively close as Sn content increases, eventually leading to a negative bandgap above 35 at.% Sn with an indirect semimetal behavior.[9] Topological Dirac semimetals could become accessible in this composition range,[10] thus enabling group IV platforms to explore exotic states of matter and potentially harness them in scalable quantum technologies.

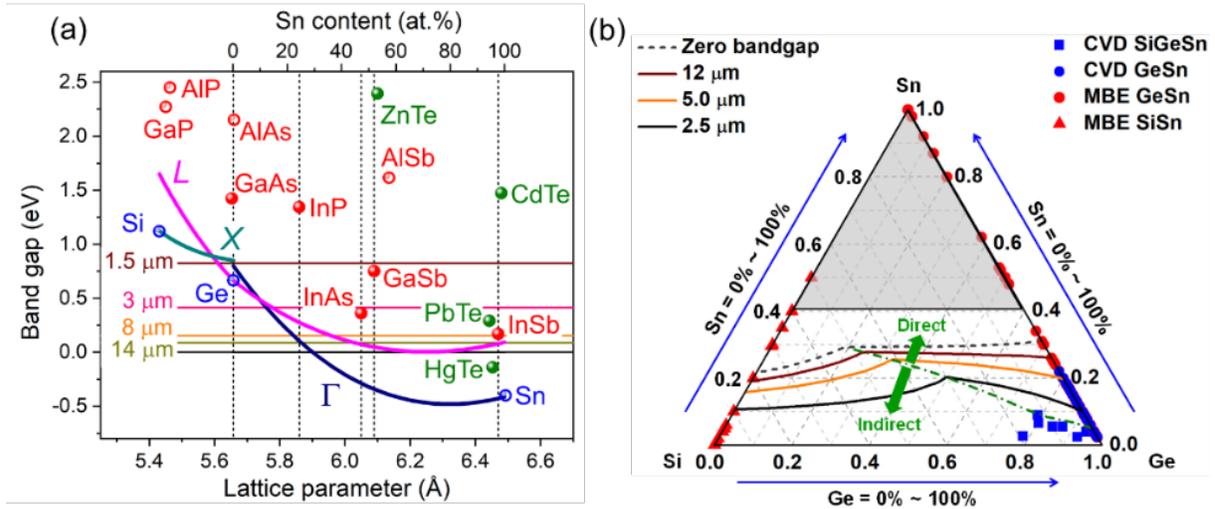

**Figure 1.** (a) Bandgap energy of GeSn as a function of the lattice parameter, and for selected III-V, II-VI semiconductors. The Sn content is shown in the top scale; (b) Composition diagram showing constant bandgap contours, indirect-direct bandgap partition, and achieved compositions by MBE and CVD methods.

Sn-containing group IV semiconductors were first mentioned in 1982 in a theoretical work that shed light on the potential of GeSn as direct bandgap semiconductors.[11] A few years later, theoretical studies supported these observations and stressed on the possibility of achieving a direct bandgap, but overestimated the Sn critical content (~20 at.%) for the indirect-to-direct band crossover.[12] However, the progress and impact of the early experimental studies were limited by multiple challenges. One of the major issues is the difficulty to obtain high-quality crystalline layers due to thermodynamic constraints limiting the solid solubility of Sn to <1 at.% in Ge and <0.1 at.% in Si, which is significantly below the critical content for the direct bandgap (≥~8 at.%). Consequently, the progress in the field required methods to achieve metastable, diamond structure (Si)GeSn materials. Recently, novel epitaxial processes using molecular beam epitaxy (MBE) or chemical vapor deposition (CVD) revived the interest in this system, allowing better control of the crystalline quality of metastable (Si)GeSn-based layers and heterostructures. Fig. 1(b) exhibits the composition range currently achieved by these two methods. Although the growth of high crystalline quality alloys at compositions across the entire pseudoternary phase diagram remains an open challenge, the last few years witnessed exciting developments at the device level. For instance, optically-pumped GeSn lasers were demonstrated at temperatures progressively approaching room-temperature using both pulsed and continuous wave (CW) excitation,[1-4] and the first electrically-pumped laser was demonstrated in 2020 at cryogenic temperatures.[5] Photodetectors covering the SWIR range and part of the MWIR range were also demonstrated.[6,7,13] Despite these promising proofs of concept, (Si)GeSn-based devices are still in their infancy and face serious challenges that need to be addressed to make them reliable technological building blocks. This Perspective highlights major hurdles in materials development, theoretical modeling and device processing, and discusses potential strategies to bring the (Si)GeSn system closer to maturity.

**Materials.** The broadly employed approach for the monolithic integration of (Si)GeSn on Si has been epitaxial growth using Ge as interlayer. Consequently, the grown layers are typically compressively strained. This compressive strain has been identified as a limiting factor preventing the incorporation of Sn during growth, and its gradual relaxation yields a composition gradient in thick layers. Plastic relaxation through the growth of lower Sn content buffer layers has been demonstrated to be effective to grow GeSn alloys with uniform Sn incorporation in the 10-20 at.% range with high crystalline quality, sharp interfaces and absence of Sn clusters.[14-16] This process is illustrated in Fig. 2. Dislocations in the GeSn multi-layer structure are mainly confined in the lower Sn content buffer layers, while high structural quality is obtained in the upper portion of the heterostructure. However, the underlying defective buffer layers can be a source of non-radiative recombination and leakage current, with a detrimental effect on device performance. Moreover, the viability of this growth process to achieve micron-thick layers with high and uniform Sn content or multilayer structures with modulated composition is yet to be demonstrated.

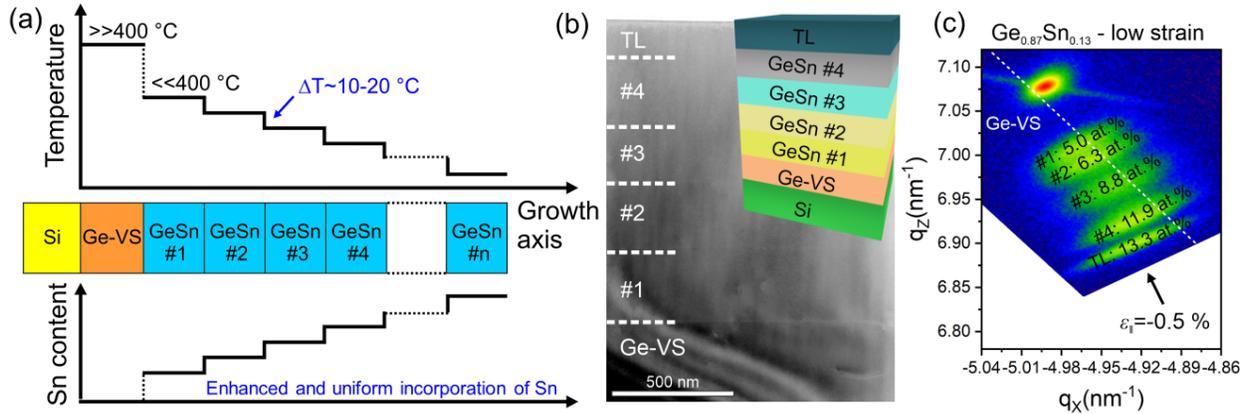

**Figure 2.** (a) Schematics of GeSn multi-layer heterostructures grown on a Si wafer. Tight control of the growth temperature yields an enhanced and uniform incorporation of Sn. (b) Cross-sectional TEM image recorded along the [110] zone axis of $Ge_{0.87}Sn_{0.13}$ grown by a step-graded process.[7] (c) The corresponding reciprocal space map around the asymmetrical (224) reflection.[7]

Despite the progress in CVD and MBE processes, alternative growth protocols and designs are still needed to obtain higher Sn content materials and also improve the crystalline quality to enhance device performance. One rather challenging but ambitious possibility would be the direct growth of GeSn on Si and/or to promote more Lomer (pure-edge 90°) which would efficiently relax the strain without creating a high-density of threading dislocations. The recently reported growth of strain-free GeSn-on-Si layers reaching a Sn content up to ~30 at.% lays the groundwork to exploit this approach and test its suitability to grow device-quality, thick layers.[17] Future improvements should also include selective-area growth and *in situ* low temperature thermal cyclic annealing processes, similar to what is currently used for the growth of Ge or SiGe on Si.[18] However, a low thermal budget must be considered to prevent any degradation of GeSn layers, which are inherently metastable. *Ex situ* treatments such as laser or flash lamp annealing can also be envisaged to reduce dislocation density. However, systematic studies to elucidate the recrystallization process of laser-annealed (Si)GeSn will be needed to identify key experimental parameters to preserve crystallinity and improve basic properties. Alternatively, the use of nanoscale growth substrates can be effective to mitigate the strain-related effects in (Si)GeSn and enhance the incorporation of Sn.[19,20] Free-standing nanomembranes made of Ge or SiGe are among many templates that can be explored. After their release and transfer onto foreign materials, they would offer highly compliant substrates partitioning the strain with the growing layer.[19] The growth of a Sn-rich layer on a nanostructured substrate would strongly reduce the compressive strain, which could result in higher Sn incorporation without segregation and phase separation. This strategy has already been shown to be effective for the growth of nanowire heterostructures.

Enhanced strain relaxation along the radial direction of the nanowire core yields defect-free, high Sn content, fully-relaxed Sn-containing shells.[21-25]

Controlled growth of lattice-matched SiGeSn/GeSn heterostructures across a broad composition range is another important challenge that would greatly improve the performance of optoelectronic devices by enhancing electron and hole confinement. Despite that the equilibrium solubility of Sn in Si is limited to ~0.1 at.%, SiGeSn layers with a tunable composition (4-20 at.% Si, 1-5 at.% Sn) and bandgap in the 0.6-1.2 eV range have been demonstrated at growth temperatures that are compatible with the growth of binary GeSn layers.[26] However, the integration of SiGeSn/GeSn heterostructures in devices is still at an early stage.[27] While an increased optical emission is expected for SiGeSn/GeSn multi-quantum well (MQW) heterostructures, the band offset must be sufficiently large to be effective for room-temperature operation. Higher Si content is therefore needed to preserve the confinement at room temperature in the GeSn active layer. In the recently demonstrated electrically-pumped GeSn laser, a 250 nm-thick p-type $Si_{0.03}Ge_{0.89}Sn_{0.08}$ layer was used.[5] Thicker SiGeSn layers (above 500 nm) would be necessary to further increase the lasing mode overlap in GeSn and reduce optical losses away from the SiGeSn contact region. Emitters will also benefit from developing n-type SiGeSn barriers to increase hole confinement in p-i-n or n-i-p structures.

As-grown intrinsic (Si)GeSn layers have an inherent background doping typically in the low $10^{17}$ $cm^{-3}$ range. This behavior is attributed to point defects (vacancy type) and to thermally generated free carriers in such narrow bandgap semiconductors. However, in-depth studies of the underlying origin of this behavior are still conspicuously missing in the literature. These studies are essential to devise effective processes to control any possible effects of growth defects on optical and electronic properties of as-grown (Si)GeSn heterostructures. The background doping combined with defective buffer layers can result in a large leakage or dark current in optoelectronic devices, thus limiting their efficiency. This can be mitigated to a certain extent by using p-i-n heterostructures. However, both n- and p-type dopant incorporation impacts the Sn content in the growing layers, which is attributed to a surface-site competition mechanism between the dopants and Sn atoms.[28,29] While these layers have been integrated in detectors and emitters, fundamental studies correlating device performance with the structural properties would shed light on the current limitations. Future works should also investigate the randomness of the doped (Si)GeSn layers down to the atomic scale to elucidate possible Sn clustering and short-range ordering and their potential impact on the optoelectronic properties.[30] The growth of high Sn content layers would also greatly benefit from the development of new precursors to provide higher reactivity, decomposition at lower temperature and higher growth rates.[31]

**Theory.** Establishing predictive theoretical models and simulations is vital to accurately design new (Si)GeSn materials and devices. Despite numerous investigations, most published theoretical calculations rely on empirical fitting to match experimental measurements. This highlights the need to consider the impact of atomistic effects on the electronic structure evolution in (Si)GeSn. The "inverted" direct bandgap of semimetallic α-Sn motivated Jenkins and Dow[12] to propose, based on tight-binding (TB) calculations employing the virtual crystal approximation (VCA), that a direct bandgap should emerge in $Ge_{1-x}Sn_x$ for $x$ > 15%. Experimental measurements confirmed the transition to a direct bandgap,[32,33] but showed the bandgap decreasing more rapidly with increasing Sn content $x$ than predicted by TB and empirical pseudopotential VCA calculations. Small supercell calculations have subsequently been undertaken by several groups to account explicitly for atomistic effects arising due to differences in size and electronegativity between the group IV elements. Strong bandgap bowing was shown by Moontragoon *et al.*[34] using 8-atom supercell charge self-consistent pseudopotential calculations, demonstrating significant deviations from equivalent VCA calculations. An accurate description of the energy gap and of the $\Gamma_{7c}$-$L_{6c}$ conduction band (CB) energy splitting in Ge can be obtained using density functional theory (DFT) based on the modified Becke Johnson (mBJ) local density approximation. DFT-mBJ calculations for 216-atom[35] and 54-atom[36] $Ge_{1-x}Sn_x$ alloy supercells have both shown strong bandgap bowing, indicative of the importance of explicitly including atomistic effects to describe the alloy band structure. Strong mixing (hybridization) is also observed between the lowest conduction state at Γ and other low-lying conduction states which originate from close to the L point in the supercells considered.[36]

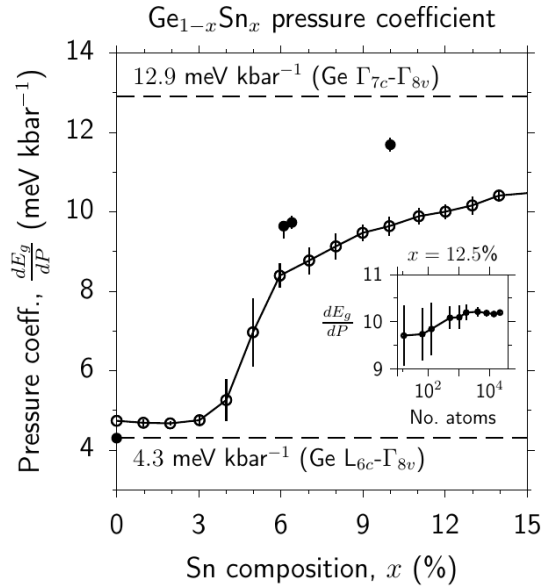

**Figure 3.** Bandgap pressure coefficient *vs.* Sn content $x$ for $Ge_{1-x}Sn_x$, measured[39] (closed circles) and calculated via TB for 1024-atom supercells (open circles). Theoretical data points represent values averaged over those calculated for 25 distinct disordered supercells; error bars represent associated standard deviations. Inset: TB-calculated pressure coefficient vs. supercell size for $x$ = 12.5%.

The strong band mixing and bandgap bowing can be understood by considering an ordered supercell, such as $Ge_{15}Sn_1$ ($x$ = 6.25%) where one Ge atom has been replaced by Sn. It can be shown for such a supercell that the lowest conduction state consists primarily of an admixture of two Ge states: (i) a linear combination $L_{6c}(A_1)\rangle$ of Ge $L_{6c}$ CB edge states having $A_1$ symmetry (i.e. purely $s$-like orbital character) at the substitutional Sn lattice site, and (ii) the zone-center Ge CB edge state $\Gamma_{7c}\rangle$. This mixing between $\Gamma$ and L-related conduction states then pushes the lowest conduction state downwards in energy, causing bandgap bowing. It is most pronounced in ordered supercells, but persists in all small $Ge_{N-M}Sn_M$ supercells ($N <\sim 200$) which can be investigated via DFT. The question then arises as to whether Sn-induced band mixing effects are an artefact of small supercell calculations, or whether they persist in large-scale atomistic calculations and in experiment, where they will have potentially significant implications for device-relevant optical and electrical properties.

To enable large-scale calculations, a multi-scale framework for multinary group-IV alloys was recently established,[37] incorporating semi-empirical valence force field (VFF) potentials and TB Hamiltonians parametrized via hybrid functional DFT calculations. The VFF and TB models, which have been benchmarked directly against DFT calculations for small (~ 100 atom) alloy supercells,[38] respectively enable atomistic structural relaxation and electronic structure calculations to be performed for alloy systems containing > $10^4$ atoms. Applying the TB method to VFF-relaxed $Ge_{1-x}Sn_x$ supercells, it was found at fixed $x$ that alloy band mixing persists, although the scale of mixing decreases with increasing supercell size (and therefore increasing disorder) up to ~2000 atoms, as discussed further below. Hydrostatic pressure provides a useful experimental technique to quantify band mixing in tetrahedrally bonded semiconductors. In Ge, this is facilitated by the very different pressure coefficients of 4.3 and 12.9 meV kbar$^{-1}$ associated respectively with the indirect (fundamental) $L_{6c}$-$\Gamma_{8v}$ and direct $\Gamma_{7c}$-$\Gamma_{8v}$ bandgaps. Given the small ($\approx$ 140 meV) $L_{6c}$-$\Gamma_{7c}$ energy difference in Ge, the supercell calculations described above suggest that Sn-induced $L_{6c}$-$\Gamma_{7c}$ hybridization should have significant impact on the nature and evolution of the $Ge_{1-x}Sn_x$ bandgap. This is supported by recent pressure-dependent spectroscopic measurements for $Ge_{1-x}Sn_x$ photodiodes.[39] The closed circles in Fig. 3 show the measured pressure dependence of the band edge in $Ge_{1-x}Sn_x$ samples, obtained from photovoltage measurements of $Ge_{1-x}Sn_x$ photodiodes as a function of hydrostatic pressure, while the open circles show the pressure dependence calculated using the TB method for a series of 1024-atom supercells. We see in both cases that the pressure coefficient for $x \sim$ 6% is intermediate between those associated with the indirect and direct bandgaps of Ge, suggesting a hybridized bandgap which is neither purely direct nor indirect in nature. The 1024-atom TB calculations underestimate the experimental measurements at higher compositions. The inset to Fig. 3 shows the calculated supercell size-dependent pressure coefficient for $x$ = 12.5%, highlighting that large-scale calculations are required to quantitatively understand electronic structure evolution in the presence of alloy band mixing effects.

Given the difficulties in extracting accurate (Si)GeSn band parameters from first principles calculations, most material and device analysis has to date been based on models parametrized empirically to treat the dependence of the bandgap on Sn composition and strain. This approach has successfully guided impressive progress to date. Strong band mixing is observed in all atomistic theoretical calculations, which persist in large supercell empirical calculations. Experimental evidence for band mixing and its consequences is to date limited. Further theoretical and experimental work is therefore now required, firstly to enable more detailed analysis of the band structure, and ultimately to enable predictive analysis of the impact of previously overlooked atomistic effects on device characteristics.

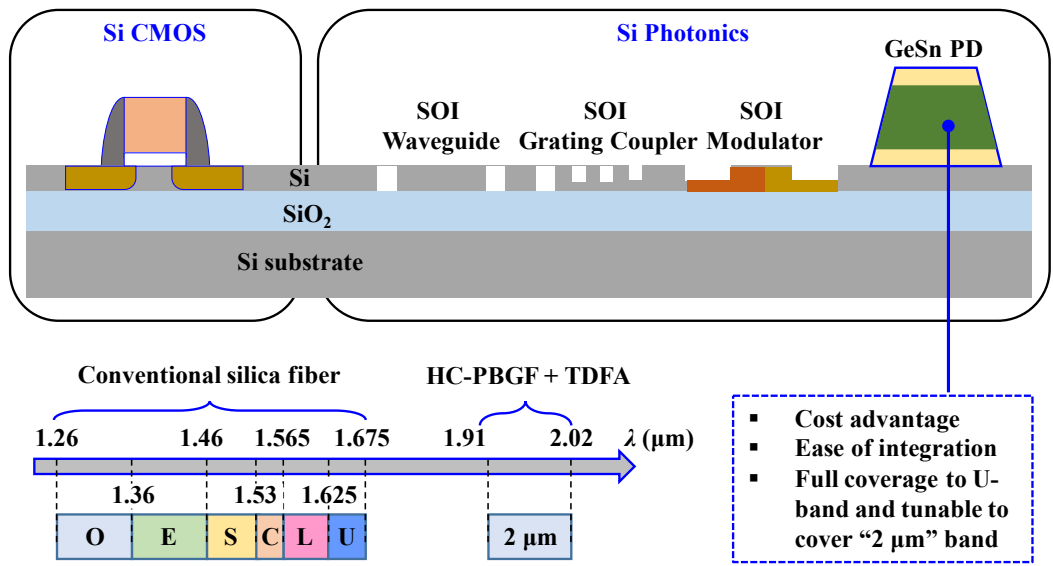

**Figure 4.** Optical communication wavelength bands using conventional silica fibers and the novel 2 μm band. Solutions using GeSn materials for photodetection and monolithically integrating them with waveguides, grating couplers, and modulators on a common SOI substrate could offer cost advantages, ease of integration with mainstream complementary metal-oxide-semiconductor (CMOS) technology, and full coverage from conventional telecommunication bands to the 2 μm band.

**Photodetectors and sensors.** The availability of high quality and stable (Si)GeSn alloys would be a disruptive technology to revolutionize SWIR, MWIR, and LWIR sensing and imaging, which can enable lighter, faster, higher signal-to-noise, integrated, and more energy efficient devices at significantly lower cost. Moreover, (Si)GeSn can also play an immediate role in extending the capabilities of Ge-based Si photonics and its flagship applications in data communications and optical components. Current Ge photodetectors only provide good coverage up to the C-band (1530-1565 nm) and their efficiency drops significantly in the L-band (1565-1625 nm).[40] Given the requirement for longer wavelength operation to support further growth in broadband data throughput, new types of photodetectors with extended detection capability are strongly needed. GeSn photodetectors can fulfil this requirement while preserving Si-compatibility. Indeed, with a low Sn content of ~2-3 at.%, full coverage of the O- to U- telecommunication bands with wavelengths from 1260 to 1675 nm can be achieved.

Higher Sn contents can expand this coverage to much longer wavelengths. Recent studies in low-loss hollow-core photonic bandgap fibers find the lowest-loss window shift from 1.55 to 2 μm, where a new and promising spectral window could be exploited for communication applications[41] to address the forecasted upcoming 'capacity crunch'. The theoretically predicted minimum loss is even lower than that of conventional single mode fibers. Thulium-doped fiber amplifiers with high gain and low noise in the two-micron-wavelength window (1910-2010 nm) further enhance its practicality to serve as a supplement to current telecommunication infrastructure, as illustrated in Fig. 4. Current photodetectors operating at 2 μm are mainly based on III-V materials such as InGaAs and GaSb, which are difficult to integrate monolithically on a Si substrate or with Si photonics. High volume manufacturing is less likely due to the limited scalability of native III-V substrates. Monolithic integration based on (Si)GeSn materials could offer cost advantages and ease of integration with mainstream complementary metal-oxide-semiconductor (CMOS) technology (Fig. 4). In this vein, a high speed GeSn p-i-n photodiode integrated on a 300 mm Si wafer was recently demonstrated at 2 μm with a 3-dB bandwidth above 10 GHz.[6] An ultra-low leakage current density of 44 mA/cm$^2$ (at a reverse bias of 1 V) was achieved, similar to that of a Ge photodiode. As compared to p-i-n photodiodes, avalanche photodiodes (APDs) can provide a significant improvement in detector responsivity due to the internal carrier multiplication mechanism. The first GeSn-on-Si APD, in which separate GeSn and Si layers are respectively employed for light absorption and carrier multiplication, was achieved with a cut-off wavelength beyond 2 μm.[42] The obtained thermal coefficient of the avalanche breakdown voltage is 0.053% K$^{-1}$, smaller than that of APDs with III-V multiplication material. The smaller thermal coefficient indicates that the GeSn-on-Si APD has a less stringent requirement for temperature stability.

Current mainstream Si photonics is built on Si-on-insulator (SOI) platforms.[43] Photonic components, including low-loss waveguides, (de)multiplexers, Mach-Zehnder interferometer modulators, and micro-ring modulators, have been realized on SOI with excellent performance at wavelengths in the 1.31-1.55 μm range. Analogously, Ge-on-insulator (GeOI) is expected to deliver similar building blocks for applications at 2 μm and beyond, as Ge becomes transparent in that wavelength range. In the short term, the introduction of GeSn into mainstream Si photonic chips, i.e., wavelength division multiplexing transceivers, could replace Ge photodetectors and extend the detection range up to the U-band to fully exploit the telecommunication bands. To achieve high responsivity and a longer detection wavelength range, high crystalline quality GeSn with higher Sn content would be needed. Currently, integrated Ge waveguide photodetectors are selectively grown on SOI photonics platform. So far, there have been no reports on GeSn waveguide photodetectors integrated on SOI. Selective area growth of GeSn on SOI and realization of SOI waveguide-based photodetectors should also be explored. The race is indeed still on to establish optimal growth processes and Sn content to reach the targeted performance while providing a reliable processing window.

With their tunable bandgap energy in the SWIR to LWIR range (Fig. 1(a)), the potential of (Si)GeSn materials extends beyond optical communications. Indeed, these systems can also be exploited to implement large-scale sensing devices for thermal imaging as well as for the spectral identification of substances sharing O-H, C-H and N-H bonds. Current SWIR and MWIR technologies are based predominantly on InGaAs, InSb, or HgCdTe. These materials are prohibitively expensive, which translates into a limited imaging and sensing array size and an overall cost of a megapixel sensor that can exceed tens of thousands of dollars. Several advantages, such as its compatibility with CMOS for large-scale manufacturing utilizing state-of-the-art Si image sensor circuits on the same chip, make (Si)GeSn a promising material system for next-generation IR sensing technologies.

As mentioned above, extended SWIR detectors based on GeSn at a Sn content up to 10 at.% have already been demonstrated. For MWIR detectors, there are limited experimental reports; however recent results show that coverage can be extended to this wavelength range.[7,15] Although the Sn content needed to cover the whole MWIR band has been achieved by several groups, the inherent compressive strain in the as-grown epi-layers limits the actual wavelength coverage. Strain engineering could be an effective strategy to get around this issue.[44-46] Regardless of the operation wavelength, adding Sn into Ge is typically associated with an increase in the dark current density as compared to that of standard Ge detectors. This is mainly due to the degradation of material quality, re-emphasizing the need to reduce or eliminate growth defects. For mature IR detector materials such as HgCdTe and InGaAs, empirical rules, for example, under the diffusion current limited conditions, have been developed to give a standard comparison of different devices at different longwave cutoffs under variable temperatures. A similar methodology should be adopted by the (Si)GeSn community to fairly compare the device performance and document the development progress, which requires reporting of the complete set of device characteristics such as temperature dependent dark current, spectrum response, responsivity, and detectivity.

For LWIR detectors, there are initial optical absorption data for both MBE- and CVD-grown GeSn that can reach this band. However, more material development effort is needed before devices can be fabricated. A possible new direction worth exploring is to evaluate if lattice-matched SiGeSn/Ge quantum wells could be used to develop L-valley inter-subband quantum well infrared photodetectors (QWIPs) for LWIR applications. The advantage of group-IV over III-V materials for QWIPs is its much longer non-polar phonon-dominated scattering lifetime in comparison with that of III-V materials that is dominated by a much shorter polar phonon scattering process.

Beyond these immediate applications, (Si)GeSn is also a testbed to address fundamental questions in photodetection. For instance, in searching for new IR detection materials it would be instructive to evaluate if indirect bandgap IR materials would offer any advantage over direct bandgap materials. Or, from a more general perspective, what is the ideal band structure for infrared materials that gives the highest detector performance? This question has been overlooked by the community due to a lack of suitable materials to study. In general, the nature of a photodetector is to generate and collect photoexcited carriers. A long carrier lifetime means more carriers to be collected. A dramatic difference of (Si)GeSn over all dominating IR semiconductors is its unique ability to be either a direct or an indirect narrow bandgap material and therefore to tune the carrier lifetime at the same optical transition energy.

| Parameter Table | HgCdTe | T2SL | InGaAs | Si | Ge | $Ge_{0.9}Sn_{0.1}$ | SiGeSn |
|---|---|---|---|---|---|---|---|
| **Bandgap** | Direct | Direct | Direct | Indirect | Quasi-direct | **Direct** | **Quasi-direct** |
| **Radiative ($cm^3/s$)** | $2\times10^{-11}$ | $1\times10^{-10}$ | $>1\times10^{-10}$ | $1\times10^{-14}$ | $6\times10^{-14}$ | **$\sim10^{-10}$** | **$\sim10^{-11}$-$10^{-13}$** |
| **Auger ($cm^6/s$)** | $2$-$3\times10^{-26}$ | $1.6\times10^{-26}$ | $10^{-29}$ | $10^{-30}$ | $10^{-30}$ | **$\sim10^{-28}$** | **$\sim10^{-28}$-$10^{-29}$** |
| **Lifetime (μs)** | ~20 | ~10-20 | ~10 | >1000 | ~1000 | **>10** | **>100** |
| **Absorption ($cm^{-1}$)** | Med. | Med., long tail | High | Low-Med. | Med.-High | **High** | **Med – High** |

**Table I. Summary of key parameters of (Si)GeSn benchmarked against other popular materials.**

Table I presents as benchmark the upper (Si)GeSn carrier lifetime limit (estimated by using experimentally reported data on Si and Ge) and theoretically calculated $Ge_{0.9}Sn_{0.1}$ radiative and Auger recombination coefficients.[47,48] When (Si)GeSn is converted from direct to indirect bandgap by introducing Si into GeSn, both radiative and Auger recombination coefficients will decrease, leading to an extended carrier lifetime on the order of hundreds of microseconds and a corresponding long diffusion length, which could potentially surpass those of HgCdTe, type II superlattices (T2SL), and InGaAs once the material quality is improved.[49] Meanwhile, for absorption, experimental results show that the GeSn absorption spectrum under direct-indirect transition has a high value of ~10,000 $cm^{-1}$ with a steep longwave cutoff[50] - twice that of HgCdTe and T2SL. Thus, it is possible to use band structure engineering to search for a sweet spot at which the material absorbs like a direct bandgap but possesses a long carrier lifetime as if it were an indirect bandgap material. In fact, when a material is under "quasi-direct" condition (i.e. direct valley is close to but slightly higher in energy than the indirect valleys), high absorption results from the direct optical transition; however, subsequently electrons could quickly scatter to the indirect valleys which could result in a long carrier lifetime due to suppressed radiative and Auger recombination. This so called "k-space charge separation" would only come into play in high quality materials in which Shockley-Read-Hall (SRH) recombination is negligible compared with radiative and Auger recombination, and its validity is yet to be confirmed.

**Light emitters**. Since the demonstration of the first optically pumped GeSn laser at 90 K, substantial progress has been made to reach higher operation temperatures, lower thresholds, CW operation, and electrical pumping.[1-5] Nevertheless, additional efforts are still needed to integrate all these characteristics in a single device. As mentioned above, improving growth protocols to achieve higher Sn incorporation is of paramount importance. This would increase the energy separation between the Γ- and L-valleys, which is an essential parameter for GeSn gain materials. A smaller Sn content results in a larger fraction of photo-excited electrons populating the L-valleys due to band-filling and thermalization effects. This increases the required pump thresholds and the device temperature, which, in turn, increases the L-valley population relative to Γ, resulting in a runoff of free carrier absorption.

Pump thresholds on the other hand can be improved by reducing the dimensionality of the gain material in GeSn/SiGeSn quantum wells (QWs).[51] However, with the 13 at.% GeSn alloy used in that work, the maximum operating temperature remained around 120 K, due to the limited directness of the bandgap. Moreover, they reported optical threshold levels, on the order of 35 kW/cm$^2$ at 20 K with a 1064-nm pump, which are an order of magnitude below that of comparable devices with bulk gain materials,[52] but remain more than one order of magnitude larger than that of III-V lasers with a comparable number (10) of QWs. A second breakthrough in regards to threshold was recently achieved by utilizing a highly tensile strained, low 5.4 at.% GeSn alloy, demonstrating a low threshold of 1.1 kW/cm$^2$ achieved in CW operation.[4] A high tensile strain of 1.4% was obtained by fully encasing the GeSn gain material in a strained silicon nitride layer, which results in a directness $\Delta E_{L-\Gamma}$ of 70 meV that is commonly only achieved at much higher Sn compositions in as-grown layers. However, the lasing temperatures being limited to 70 K and 100 K for CW and pulsed operation, respectively, indicates that GeSn with a larger directness must be reached to unlock device operation approaching room-temperature.

Dark recombination rates are likely to play a very important role in these metastable materials. Indeed, low lifetimes of 214 ps[53] have been recently reported for GeSn at Sn content x = 12.5 at.%, suggesting short nonradiative lifetimes when substantial levels of Sn are being incorporated. Since Sn incorporation depends on the growth temperature, with lower Sn content materials grown at higher temperatures, it appears likely that nonradiative lifetimes due to SRH recombination depend on the Sn content and worsen as the latter increases.[54] Another constraint in the design of GeSn/SiGeSn heterostructures is the higher growth temperature that is typically required for the SiGeSn cladding and barrier layers. In fact, low temperature Si incorporation is hindered by limited precursor (e.g., disilane) cracking. With its wider bandgap, and thus smaller refractive index, SiGeSn can serve to confine both free carriers and photons and is thus essential to improve the performance and the operating temperature of GeSn lasers. In order for MQWs involving the SiGeSn material system to operate effectively at room temperature, not only does the directness of the gain material need to be high enough, the barriers confining carriers in the QWs also need to be high enough to adequately confine thermalized carriers inside the wells. Thus, increasing the Si content in auxiliary layers is also of high importance to reach high-efficiency room temperature lasing.

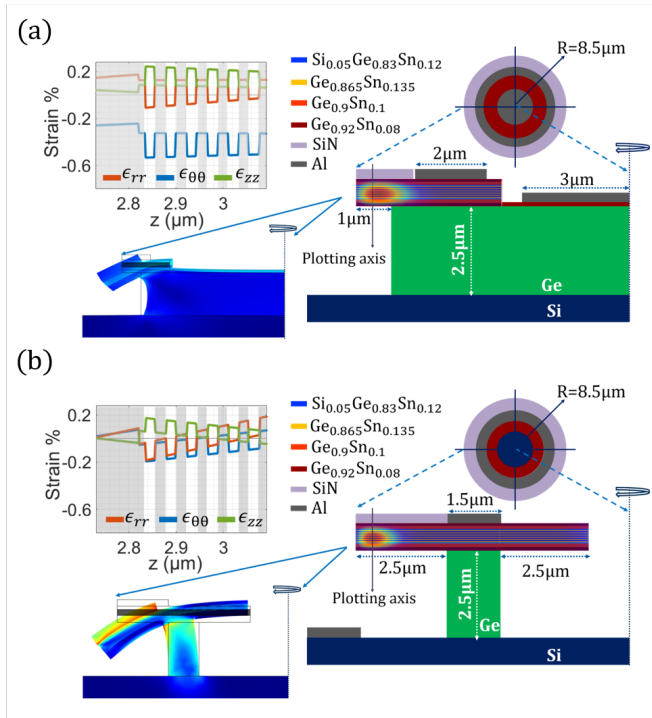

**Figure 5. Two concepts for undercut MD lasers with improved heat sinking.** The insets show the strain profile along the vertical axis drawn close to the periphery of the disk as well as an exaggerated representation of the deformation occurring after undercut. Modeling is based on a 752 nm thick SiGeSn/GeSn gain material stack with six 13.5 at.% GeSn QWs, with a dopant profile optimized for electrical injection and grown on top of a 2.5 $\mu$m thick virtual substrate. A 300 nm thick silicon nitride stressor layer, with an initial compressive hydrostatic stress of -1 GPa and -2.8 GPa, respectively for panels (a) and (b), deforms the membrane. In addition to radial extension, it bows slightly downward at its outer periphery (23.6 nm in (a), 126.7 nm in (b)). In both cases, the residual radial strain component drops to almost zero; however, in (a) the azimuthal compressive strain remains at -0.5%, close to its as-grown value. In (b), the inner part of the annulus shaped membrane is also under etched to allow further deformation, with the pillar bending outwards and the membrane bending further down at its outer edge. Residual compressive strain in the wells is close to biaxial and remains below -0.18%, close to what was achieved for the optically pumped structure reported by D. Stange *et al*.[52]

As mentioned above, externally induced tensile strain provides an additional means to improve the directness without sacrificing material quality or the Si content of barrier layers. In the case of electrically pumped structures, free standing membranes, however, present an additional level of complexity, as they increase electrical path lengths and worsen heat sinking. Similar difficulties were for example encountered with electrically pumped III-V photonic crystal membrane lasers, which were solved with a pedestal placed right below the nanocavity serving as both a heat sink and an ideally placed electrical connection.[55] Fig. 5(a) shows a concept for an undercut microdisk (MD) laser, in which the underlying Ge is etched away to a much smaller extent to provide better heat sinking while allowing for membrane deformation in the optically active region. A silicon nitride stressor, deposited and patterned on top of the SiGeSn gain stack, induces tensile stress resulting in radial extension in the undercut region ($\epsilon_{rr}$). Note that the strain in the azimuthal direction, $\epsilon_{\theta\theta}$, remains very close to that of the initially grown structure. The reduction of the hydrostatic compressive strain (as averaged over all three spatial dimensions), or conversion of the hydrostatic strain component from compressive to tensile, improves the directness of the material and thus enables higher temperature lasing.

It is worth noting that biaxial compressive (tensile) strain in the previously described structures increases the gain of modes with strong in-plane (out-of-plane) E-field components including quasi-transverse-electric (TE) polarized (quasi-transverse-magnetic (TM) polarized) whispering gallery modes (WGMs). However, the uniaxial compressive strain along the azimuthal direction seen here favors modes with E-fields oriented along the azimuthal direction, which correspond to radially propagating Fabry-Perot resonances, that are not supported by this structure. In order to remedy this problem, annulus-shaped structures can be used (Fig. 5(b)). By replacing the central pedestal with a ring-shaped structure, that is etched from both sides, higher deformation of the structure is achieved, with the top of the pedestal bending outwards and a more pronounced bending of the outer membrane edge towards the substrate (insets show an exaggerated representation of the deformation). This allows to relieve the compressive strain in both in-plane directions. Modeling of these structures at 100 K indicates that these selection rules might play less of a role then first expected, as a consequence of population of both heavy- and light-hole bands as well as polarization-dependent inter-valence band absorption. Thus, while the uniaxial compressive strain seen along the azimuthal direction is detrimental to the amplification of WGMs, and the structure in Fig. 5(b) is preferable from that perspective, it might not prevent the structure in Fig. 5(a) from lasing and constitutes only one of several considerations in the device optimization. Concepts developed for waveguide coupling MD lasers[56] also have yet to be adapted to undercut structures.

In both designs in Figs. 5(a) and 5(b), the top contact is drawn inside the disk, away from the outer periphery at which the lowest order WGMs have their maximal field strength, in order to reduce resulting optical absorption losses. This type of contacting scheme works very well for III-V disk lasers heterogeneously integrated into silicon photonics,[57] in which, above threshold, the increased radiative recombination in the high optical field regions leads to effective diffusion-driven carrier transport to the active gain medium. However, low SRH lifetimes with current SiGeSn materials would reduce the efficiency of this carrier transport mechanism. Combination of vertical light confinement by means of wider bandgap optical cladding layers, enabling top electrode placement right above the optically active gain region while maintaining acceptable optical losses, with the strain engineering proposed here, might be conducive to obtain even better performance electrically pumped lasers.

In addition to lasers, (Si)GeSn light-emitting diodes (LEDs) should also be properly investigated because of their relevance as sources of free-space incoherent optical emission needed in biomedical and gas sensing applications.[58,59] The aspects of electrical injection and carrier confinement are somewhat similar to electrically pumped lasers, while the integration into an optical cavity is not required. Being spatially multimode, LEDs are not appropriate for efficient coupling with few-mode optical waveguides. The outcoupling of the incoherent emission from high index semiconductors into free space brings with it many issues, but these problems are rather well understood today. Until now, the reported (Si)GeSn LEDs were just used as test vehicles for laser experiments, with little effort spent on improving the LED itself. Surprisingly, the efficiency of the first, non-optimized GeSn LEDs was on par with commercial III-V InAsSb mid-IR LEDs.[60] A simple gas detection using GeSn LEDs was recently demonstrated.[61]

In both GeSn homojunctions and MQW heterostructures, the LED efficiency follows the same trend observed in optically pumped GeSn lasers. The best barrier material to date for QWs is the SiGeSn alloy. The barrier energy offset between SiGeSn and GeSn confines carriers in a smaller GeSn well, and strongly increases the radiative recombination. This was proved experimentally, however, for indirect GeSn.[62,63] Indeed, quantization decreases the GeSn alloy directness and lowers the barrier-well energy band offset, thus reducing the emission efficiency and limiting the confinement effect at higher temperatures. To date, the efficiency of the MQW LEDs at elevated temperatures drops well under that of bulk GeSn LEDs that employ higher Sn content, even though the higher Sn content GeSn suffers more lattice defects. More research is required to improve the epitaxy of (Si)GeSn alloys at a larger Si molar fraction in barriers to extend the MQW advantages towards room temperature operation of LEDs.

**Summary.** This Perspective provides a glimpse into the development and potential of (Si)GeSn semiconductors and how they brought to the limelight an ambitious roadmap to achieve the long-sought-after monolithic integration of electronics and photonics. The ability to precisely and simultaneously control lattice parameters and electronic bands in these group IV alloys enables a variety of heterostructures and low-dimensional systems, which are versatile building blocks for silicon-compatible infrared detectors, sensors, and emitters. Despite recent major advances, further hurdles face the development of these technologies. From our analysis, overcoming thermodynamic constraints to develop high-quality materials remains the most pressing challenge that needs to be solved. The collective efforts of the community are necessary to improve growth processes and identify key experimental parameters to achieve optimal (Si)GeSn layers and heterostructures for higher device performance. The race towards reliable (Si)GeSn technologies would also benefit from predictive theoretical models and simulations and adapted device processing protocols. Current device performance falls short when compared with that demonstrated by established compound semiconductors. Nevertheless, one should not ignore that the latter have already benefited from decades of research and development.[64,65] We envision that similar efforts will be crucial to advance (Si)GeSn materials and harness their properties in innovative silicon-integrated infrared technologies.

**Acknowledgements.** O. Moutanabbir acknowledges the support from NSERC Canada (Discovery, SPG, and CRD Grants), Canada Research Chairs, Canada Foundation for Innovation, Mitacs, PRIMA Québec, and Defence Canada (Innovation for Defence Excellence and Security, IDEaS). E. O'Reilly and C. Broderick acknowledge the support of Science Foundation Ireland (SFI; project no. 15/IA/3082), and the National University of Ireland (Post-Doctoral Fellowship in the Sciences to C.B.). D. Buca, J. Witzens and B. Marzban gratefully acknowledge funding from the "Deutsche Forschungsgemeinschaft" for grant no. 299480227. D. Nam acknowledges the support from National Research Foundation of Singapore through the Competitive Research Program (NRF-CRP19-2017-01). S.-Q. Yu acknowledges the financial support from the Air Force Office of Scientific Research (AFOSR) (Grant Nos. FA9550-18-1-0045, FA9550-19-1-0341). A. Chelnokov acknowledges AAPG and Carnot grants by the French ANR.

The following article has been submitted to Applied Physics Letters. After it is published, it will be found at https://aip.scitation.org/journal/apl.

**Data availability.** The data that support the findings of this study are available within this article.